\documentclass[conference,10p]{IEEEtran}
\let\labelindent\relax

\usepackage[normalem]{ulem}
\usepackage{enumitem}
\usepackage{slashbox}
\usepackage{ifpdf}
\usepackage{multirow}
\usepackage[hyphens]{url}
\usepackage{graphicx}
\usepackage{amssymb}
\usepackage{amsmath}
\usepackage{color}
\usepackage{tabu}
\usepackage{lipsum}% http://ctan.org/pkg/lipsum

\usepackage{graphicx}
\usepackage{caption}
\usepackage{subcaption}
\usepackage{cite}

\ifCLASSINFOpdf

\else
	
\fi

\begin{document}

\title{Investigation of Performance in Integrated Access and Backhaul Networks}

\author{
\IEEEauthorblockN{Muhammad Nazmul Islam, Navid Abedini, Georg Hampel, Sundar Subramanian and Junyi Li}
\IEEEauthorblockA{Qualcomm Flarion Technologies, Bridgewater, New Jersey }}

\maketitle

\begin{abstract}

Wireless backhaul allows densification of mobile
networks without incurring additional fiber deployment cost.
This, in turn, leads to high spatial reuse, which is a significant tool
to meet increasing wireless demand in 5G networks.
Integrated access and backhaul (IAB), where access and backhaul network share
the same standard wireless technology (e.g. 5G new radio (NR) standard),
allows interoperability among different IAB manufacturers and
flexible operation between access and backhaul.
This paper investigates joint resource allocation and relay selection 
in a multi-hop IAB network to maximize geometric mean of UE rates. 
Our study illustrates several advantages and features of IAB.  
First, IAB significantly improves UE rates compared to access only networks and
can provide an important intermediate solution during incremental fiber deployment.
Second, IAB networks with optimal mesh outperforms IAB networks
with RSRP based spanning tree both in terms of rate and latency.

\end{abstract}


\section{Introduction}

Wireless demand is expected to increase rapidly over the next few years. Higher
bandwidth and spatial reuse are necessary to meet this increasing demand~\cite{Andrews}.
Due to the availability of abundant bandwidth and higher spatial reuse through directional beamforming,
millimeter wave (MMW) bands ($10$ times the current carrier frequency of $3$ GHz) can fulfill both of these criteria.
That is why, millimeter wave based cellular access has been an integral part of LTE Rel-15~\cite{Rel-15}.  
									
Signals transmitted via millimeter wave band, however, suffer from high path loss due to the use of higher carrier frequency and different 
additional propagation losses like oxygen absorption loss, rain absorption loss, etc.~\cite{Farooq}.
Hence, millimeter wave based cellular access is only feasible for small cell networks. Providing wired backhaul to many small cells may
dramatically increase fiber deployment cost. Wireless backhaul allows network operators to flexibly deploy small cell base stations without incurring additional fiber deployment cost. Wireless backhaul is also an important tool during \emph{incremental
deployment} of fiber in mobile networks. During the early stages of a network rollout,
fiber can be deployed to a subset of base stations, also known as anchor nodes, and the access traffic
of the remaining base stations can be wirelessly backhauled to the anchor nodes.																						
As network traffic demand grows, fiber can be deployed to all base stations of the network to further enhance capacity.

Wireless backhaul can be implemented using different techniques. Integrated 
access and backhaul (IAB), where access and backhaul communications use the same 
standard radio technology (e.g. 5G NR), allows interoperability among 
base stations from different manufacturers, which is essential
for flexible deployment of dense small cell networks~\cite{Navid}. IAB can be deployed through both
in-band and out-band relaying and used in both indoor and outdoor networks. 
This paper investigates joint resource allocation and relay selection 
in a multi-hop IAB network to maximize geometric mean of UE rates. 
Our study illustrates several advantages and features of IAB.  
%Fig.~\ref{fig:IAB_Use} shows
%various use cases of an IAB network.

\subsection{Related Work}

Several works have focused on resource allocation, relay selection and fiber deployment
in wireless backhaul networks. 
The authors of~\cite{Ismail,Ehsan,Nazmul1,Shen} investigated optimal
relay placement in backhaul networks to meet a certain demand at base stations. 
Rasekh et. al.~\cite{Madhow} tackled the issue from the other direction, i.e., they 
performed joint resource allocation to maximize rate in a wireless backhaul network with a fixed set of anchor nodes.

The authors of~\cite{Nazmul2} and~\cite{Andrews2} extended above works to IAB networks. Islam et. al.~\cite{Nazmul2} performed joint optimal 
resource allocation, relay selection and fiber drop deployment to minimize the fiber deployment
cost while meeting UE demand. It, however, did not illustrate
UE rate distribution in variaous scenarios with fixed fiber deployment in the networks.
Kulkarni et. al.~\cite{Andrews2}
focused on an IAB based millimeter wave cellular network
and investigated the performance of dynamic TDD networks with unsynchronized and access-backhaul split
at different base stations. It constrained the maximum number of hops between UE
and anchor nodes to be two and used Poisson point process based deployment of users and base stations to 
illustrate the rate distribution in IAB networks.

Compared to the above mentioned works, our work provides several novel findings.
First, we use distribution of UE rates to show the advantages of IAB over purely access networks in a realistic
multi-hop network setting of downtown Manhattan where the link gain between different nodes
are obtained from a ray tracing tool. Second, we show that IAB can provide an important intermediate solution 
to increase UE rates during incremental fiber deployment. Third, we illustrate how the performance of an IAB network
generated with signal-strength-based spanning tree topology compares with that of an IAB network generated with 
load-balanced mesh topology.

This paper is organized as follows. 
Section~\ref{sec:ProblemFormulation} formulates the
general optimization problem. Section~\ref{sec:DiffVersion} and~\ref{sec:SolMethod}
describe different variants of the general optimization problem and the solution methodology respectively. After providing simulation 
results in section~\ref{sec:NumericalSimulations}, we conclude our work in section~\ref{sec:Conclusion}.

Throughout this paper, the following terms may be used
interchangeably, (``base station", ``BS" and ``gNB' ), (``user equipment" and ``UE"), (``gNBs with fiber drop" and ``anchor nodes").

\section{Problem Formulation}  \label{sec:ProblemFormulation}

\begin{figure}[!t]
\centering
\includegraphics[width=3.5in]{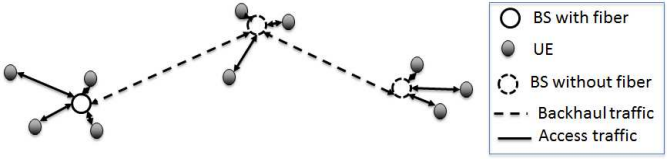}
\caption{A generic figure of integrated access backhaul network. Figure reproduced from~\cite{Nazmul2}}
\label{fig:Generic_Figure}
\end{figure}

\begin{table}
\begin{center}
\begin{tabular}{|l|l|} \hline
Parameter & Notation \\ \hline
% Set of base stations & \mathcal{N} \\ \hline
% Fiber drop deployment cost in BS $i$ & $p_i$  \\ \hline
% fber drop decision in BS $i$ & $y_i$ \\ \hline
%Set of UEs connected to BS $i$ & $\mathcal{UE}_i$ \\ \hline
%$k$-th UE connected to BS $i$ & $i_k$ \\ \hline
Set of user equipments & $\mathcal{UE}$ \\ \hline
Set of base stations & $\mathcal{BS}$ \\ \hline
Total number of UEs & $N$ \\ \hline
Capacity of the fiber pipe & $M$ \\ \hline 
\emph{Allotted time in link $ij$ for UL traffic} & \emph{$t_{ij}^U$} \\ \hline
\emph{Allotted time in link $ij$ for DL traffic} & \emph{$t_{ij}^D$} \\ \hline
%Allotted time in access link $i,i_k$ for DL traffic & $t_{i,i_k}$ \\ \hline
%Allotted time in access link $i_k,i$ for UL traffic & $t_{i_k,i}$ \\ \hline
\emph{Flow in link $ij$ for UL traffic} & \emph{$f_{ij}^U$} \\ \hline
\emph{Flow in link $ij$ for DL traffic} & \emph{$f_{ij}^D$} \\ \hline
Capacity of link $ij$ & $c_{ij}$ \\ \hline
Available connectivity of access link $ij$ & $a_{ij}$ \\ \hline
Available connectivity of backhaul link $ij$ & $b_{ij}$ \\ \hline
fiber drop decision at BS $j$ & $y_j$ \\ \hline
\end{tabular}
\end{center}
\caption{List of Notations} \label{tab:Notations}
\end{table}

%Let $\mathcal{UE}$ and $\mathcal{BS}$ denote the set of UEs and base stations in the network. Assume $|\mathcal{UE}| = N$. 

Fig.~\ref{fig:Generic_Figure} shows a generic integrated access backhaul network. Three base stations are connected to a group of UEs. 
Fiber is deployed in one base station. The other two base stations route their DL (UL) data
from (to) the anchor node through one-hop and two-hop wireless backhaul paths. 

Similar to this generic figure, we focus on a network with a
set of UEs $\mathcal{UE}$ and a set of base stations $\mathcal{BS}$. 
Our work maximizes the geometric mean (GM) of UE rates with a fixed set of fiber drops among the base stations.

Table~\ref{tab:Notations} shows the list of parameters and the corresponding
notations that we use throughout the paper. The variables are displayed with italic style in the table. 
%Note that, both $y_j  \in \{0, 1\}$ 
%and $I_{i,j} \in \{0, 1\}$ and are inputs to the optimization problem.
%$y_j$ is used throughout all the optimization problem formulations.
%$I_{i,j}$ are used to describe the effects of load balancing in access networks
%and the performance of IAB networks with RSRP based spanning tree.

We now develop our optimization problem formulation
by going through the set of constraints and optimization objective. 

\subsection{Objective}

Let $f_{ij}^{D}$ and $f_{ij}^{U}$ denote the flow between node $i$ and $j$
to carry downlink (DL) and uplink (UL) traffic respectively.
The geometric mean of all UEs' rates can be expressed as:
\begin{equation}
\bigl(\prod_{i \in \mathcal{UE}} \bigl(\sum_{j \in \mathcal{BS}} f_{ij}^{U} \bigr) \bigl(\sum_{j \in \mathcal{BS}} f_{ji}^{D} \bigr) \bigr)^{\frac{1}{2N}}
\end{equation}
The objective of the optimization problems considered in this paper is to maximize the 
above function.

 \subsection{ Flow Capacity Constraint}

%Let $d_{i_k}^D$ and $d_{i_k}^U$ denote the DL and UL demand of UE $i_k$
%respectively. The DL (UL) flow to (from) each 
%UE should exceed its DL (UL) demand. Hence,
%%
%\begin{equation}
%f_{i_k,i} \geq d_{i_k}^U \, , \, f_{i,i_k} \geq d_{i_k}^D \, 
%\forall i_k \in \mathcal{UE}_i \, , \, \forall i \in \mathcal{N} 
%\end{equation}
%%

Throughout this work, we assume that UEs always transmit at maximum
power and the power spectral density does not depend on the allocated
time duration. Hence, the flow of each link should be upper bounded by
the product of the allocated time slots and the capacity of the link.

However, we also consider available connectivity pattern $a_{ij}$ and $b_{kl}$ 
for access link $ij$ and backhaul link $kl$ respectively. These available connectivity patterns are inputs 
to the optimization problem and determine whether the particular link can be activated or not.
The specific values of these available connectivity patterns for different versions of the optimization problems are described in
section~\ref{sec:DiffVersion}. Hence,
\begin{equation}
f_{ij}^{U} \leq t_{ij}^{U} c_{ij} a_{ij} \, , \, \, f_{ji}^{D} \leq t_{ji}^{D} c_{ji} a_{ji} \, \forall i \in \mathcal{UE}, \, \forall j \in \mathcal{BS},
\end{equation}
\begin{equation}
f_{ij}^{U} \leq t_{ij}^{U} c_{ij} b_{ij} \, , \, f_{ij}^{D} \leq t_{ij}^{D} c_{ij} b_{ij} \, \forall i \in \mathcal{BS}, \, \forall j \in \mathcal{BS}, \, i \neq j.
\end{equation}
Note that, DL (UL) traffic in backhaul links refers to the portion of traffic that got generated
from DL (UL) traffic in access links.

\subsection{Flow Conservation Constraint}

We set up the flow conservation constraint model in the same way as that of~\cite{Nazmul2}.
If a BS is not connected to fiber, outgoing and incoming flow should be equal.
If a BS $i$ is connected to fiber, summation of outgoing DL traffic in both access and backhaul
should be less that the capacity of the fiber pipe at BS $i$.
Similarly, summation of incoming UL traffic in both access and backhaul should not exceed the
capacity of the fiber pipe at BS $i$.
These relations can be expressed through the following
constraints:
\begin{equation}
\sum_{l \in \mathcal{UE}} f_{il}^D + 
\sum_{j \in \mathcal{BS}, j \neq i} f_{ij}^D =
\sum_{k \in \mathcal{BS}, k \neq i} f_{ki}^D + M_i^D 
\, \, \forall i \in \mathcal{BS},
\end{equation}
\begin{equation}
\sum_{l \in \mathcal{UE}} f_{li}^U +
\sum_{k \in \mathcal{BS}, k \neq i} f_{ki}^U =
\sum_{j \in \mathcal{BS}, j \neq i} f_{ij}^U + M_i^U 
\, \, \forall i \in \mathcal{BS}.
\end{equation}
\begin{equation}
M_i^D + M_i^U \leq M y_i \, \forall \, i \in \mathcal{BS}.
\end{equation}
where $M$ and $y_i$, inputs to the optimization problem,
denote the total capacity of the fiber pipe and the fiber deployment decision at BS $i$ respectively.
$M_i^D$ and $M_i^U$ denote how BS $i$ splits
its total fiber capacity to carry DL and UL traffic respectively.

\subsection{Resource Allocation Constraint}

We assume time division multiplexing among adjacent links throughout this work
and focus on in-band relaying where the partition between
access and backhaul resources can vary across gNBs. Hence,
\begin{equation}
\sum_{i \in \mathcal{UE}} \bigl(t_{ij}^{D} + t_{ji}^U \bigr) 
+ \sum_{k \in \mathcal{BS}} \bigl(t_{kj}^{D} + t_{kj}^{U} + t_{jk}^{D} + t_{jk}^U \bigr) \leq 1
\, \forall j \in \mathcal{BS}, \, j \neq k
\end{equation}
Our work can be easily extended to frequency division multiplexing based networks
and out-band relaying techniques.

\subsection{Interference Consideration}

This paper focuses on IAB networks that operate at MMW band.
Due to the directional transmission at MMW band,
we do not assume interference among non-adjacent links
and consider interference based IAB design as a future work.
Our previous work~\cite{Nazmul2} shows that pairwise interference
from backhaul to access links
exceed noise in only less than $7\%$ cases in the network setting that we consider.
\subsection{Overall Optimization Problem}

\begin{subequations}
\begin{equation}
\max \bigl(\prod_{i \in \mathcal{UE}} \bigl(\sum_{j \in \mathcal{BS}} f_{ij}^{U} \bigr) \bigl(\sum_{j \in \mathcal{BS}} f_{ji}^{D} \bigr) \bigr)^{\frac{1}{2N}}
\label{eq:Objective}
\end{equation}
\begin{equation}
f_{ij}^{U} \leq t_{ij}^{U} c_{ij} a_{ij} \, , \, \, f_{ji}^{D} \leq t_{ji}^{D} c_{ji} a_{ji} \, \forall i \in \mathcal{UE}, \, \forall j \in \mathcal{BS}
\label{eq:FlowCapAC}
\end{equation}
\begin{equation}
f_{ij}^{U} \leq t_{ij}^{U} c_{ij} b_{ij} \, , \, f_{ij}^{D} \leq t_{ij}^{D} c_{ij} b_{ij} \, \forall i \in \mathcal{BS}, \, \forall j \in \mathcal{BS}, \, i \neq j
\label{eq:FlowCapBH}
\end{equation}
\begin{equation}
\sum_{l \in \mathcal{UE}} f_{il}^D + 
\sum_{j \in \mathcal{BS}, j \neq i} f_{ij}^D =
\sum_{k \in \mathcal{BS}, k \neq i} f_{ki}^D + M_i^D 
\, , \, \forall i \in \mathcal{BS}
\label{eq:FlowConDL}
\end{equation}
\begin{equation}
\sum_{l \in \mathcal{UE}} f_{li}^U +
\sum_{k \in \mathcal{BS}, k \neq i} f_{ki}^U =
\sum_{j \in \mathcal{BS}, j \neq i} f_{ij}^U + M_i^U 
\, , \, \forall i \in \mathcal{BS}
\label{eq:FlowConUL}
\end{equation}
\begin{equation}
M_i^D + M_i^U \leq M y_i \, \forall \, i \in \mathcal{BS}
\end{equation}
\begin{equation}
\sum_{i \in \mathcal{UE}} \bigl(t_{ij}^{D} + t_{ji}^U \bigr) 
+ \sum_{k \in \mathcal{BS}} \bigl(t_{kj}^{D} + t_{kj}^{U} + t_{jk}^{D} + t_{jk}^U \bigr) \leq 1
\, \forall j \in \mathcal{BS}, \, j \neq k
\label{eq:ResourceCon}
\end{equation}
\begin{equation}
f_{ij}^U \geq 0, \, f_{ji}^D \geq 0, \, t_{ij}^U \geq 0, \, t_{ji}^D \geq 0 \,
\forall i \in \mathcal{UE}, \, \forall j \in \mathcal{BS},
\label{eq:VariableCon1}
\end{equation}
\begin{equation}
f_{ij}^U \geq 0, \, f_{ij}^D \geq 0, \, t_{ij}^U \geq 0, \, t_{ij}^D \geq 0 \,
\forall i \in \mathcal{BS}, \, \forall j \in \mathcal{BS}, \, i \neq j
\label{eq:VariableCon2}
\end{equation}
\begin{equation}
M_j^D \geq 0, M_j^U \geq 0 \forall j \in \mathcal{BS}
\label{eq:VariableCon3}
\end{equation}
\label{eq:MainOpt}
\end{subequations}

We investigate different variants of the optimization problem mentioned
above to see the impact of IAB networks. The next section
describes these different variants.

\section{Different variants of Optimization problem}   \label{sec:DiffVersion}
\subsection{Access only, signal-strength-based cell selection}  \label{sec:AccessWithoutLB}
This scenario focuses on access only network, i.e., it does not allow any backhaul traffic.
Hence, $b_{ij} = 0 \, \, \forall i \in \mathcal{BS}, \, j \in \mathcal{BS}, \, j \neq i$.
UEs are connected to the BS based on maximum signal strength.
The possible access connectivity pattern is,
\begin{equation}
\forall i \in \mathcal{UE}, a_{ij} = 1 \, \, \emph{iff} \, \, c_{ij} \geq c_{ik} \, \forall k \in \mathcal{BS}, \, k \neq j
\end{equation}
Each UE is connected to only one BS. If, for a UE $i$, $c_{ij} = c_{ik} \geq c_{il} \, \forall l \in \mathcal{BS},\, l \neq j, l \neq k$,
then one of $a_{ij}$ and $a_{il}$ links gets randomly activated and the other remains inactive.
\subsection{Access only, load-balanced cell selection}  \label{sec:AccessWithLB}
This scenario focuses on access only network, too. Hence, $b_{ij} = 0 \, \, \forall i \in \mathcal{BS}, \, j \in \mathcal{BS}, \, j \neq i$..
However, It allows any UE to be connected to any BS. Hence, $a_{ij} = 1 \forall i \in \mathcal{UE}, \, j \in \mathcal{BS}$.
One UE could be connected to multiple BSs in this scenario.
\subsection{IAB, signal-strength-based cell selection and spanning tree (ST) topology} \label{sec:BackhaulSpanningTree}
UEs are connected to BS based on maximum signal strength. 
Hence, the access connectivity is same as that of~\ref{sec:AccessWithoutLB}.
The backhaul connectivity pattern is generated using signal strength based spanning tree.
The algorithm to generate the backhaul connectivity pattern can be described as follows:
\begin{enumerate}
\item Initialize the gNBs with fiber as connected nodes and gNBs without fiber 
as unconnected nodes.
\item \label{item:WhileLoop} Iterate through following steps till the set of unconnected nodes is an empty set.  
\item Generate edge graph between the elements of the connected node set and those 
of the unconnected node set.
\item Pick the edge with the strongest link gain. Remove the corresponding unconnected node
from the set of unconnected nodes and add it to the set of connected nodes. Go to step~\ref{item:WhileLoop}.
\end{enumerate}
\subsection{IAB, signal-strength-based cell selection and load-balanced mesh topology} \label{sec:IABWithoutLB}
UEs are connected to BS based on maximum signal strength. 
The access connectivity of this scenario is same as that of~\ref{sec:AccessWithoutLB}.
However, this scenario allows any base station to be connected with any other BS. 
Hence, $b_{ij} = 1 \forall i \in \mathcal{BS}, \, j \in \mathcal{BS}, \, j \neq i$.
\subsection{IAB, load-balanced cell selection and mesh topology} \label{sec:IABWithLB}
This scenario allows any UE to be connected to any BS and any base station to be connected with any other BS. 
Hence, $a_{ij} = 1 \forall i \in \mathcal{UE}, \, j \in \mathcal{BS}$ and
$b_{ij} = 1 \forall i \in \mathcal{BS}, \, j \in \mathcal{BS}, \, j \neq i$.
One UE could be connected to multiple BSs in this scenario.

\section{Numerical Simulations}  \label{sec:NumericalSimulations}

\begin{figure}[!t]
\centering
\includegraphics[width=2.1in]{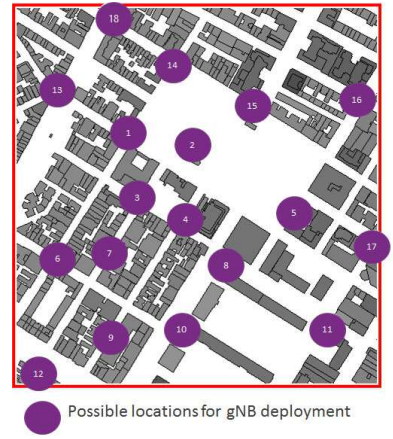}
\caption{A wireless network with 18 possible locations of base stations in downtown Manhattan. UEs are randomly distributed in streets and not shown in the figure }
\label{fig:BS_Location_Without_BS}
\end{figure}
\begin{table}
\begin{center}
\begin{tabular}{|l|l|} \hline
Parameter & Value \\ \hline
BS and UE Tx Power (dBm) & $30$ \\ \hline
BS antenna array configuration & $16 \times 8$  \\ \hline
Array gain (dB) & $21$ \\ \hline
BS EIRP (dBm) & $51$ \\ \hline
Total bandwidth (GHz) & $1$ \\ \hline
Frequency (GHz) & $28$ \\ \hline
Atmospheric absorption (dB/km) & $0.11$ \\ \hline
Polarization loss (dB) & $1$ \\ \hline
Alignment error (dB) & $5$ \\ \hline
Implementation loss (dB) & $5$ \\ \hline
Capacity of fiber pipe (Gbps) & $200$ \\ \hline
\end{tabular}
\end{center}
\caption{Link Budget Calculation} \label{tab:LinkBudgetCal}
\end{table}
We consider a wireless network setting in downtown Manhattan and use it to show the usefulness of IAB networks. 
Fig.~\ref{fig:BS_Location_Without_BS} shows the network location. The eighteen red circles denote possible locations for
gNB deployment.
The average distance between these possible sites is $200$ m. 
$600$ UEs are randomly thrown into the open areas of the grid. The link gains between
possible gNBs and UEs are obtained from WINPROP, a ray tracing tool~\cite{WINPROP}. 

We assume that each gNB has a rectangular planar array with $16 \times 8$ antenna elements.
Each UE has only one antenna element. The transmit power at both gNB and UE is assumed to be $30$ dBm.
Our work can be easily extended to scenarios where BS and UE have different transmit powers.
gNB uses constant phase offset beams and directs it towards the strongest cluster of the angle of 
arrival or departure of a particular link to communicate with a UE.
The effective signal to noise ratio (SNR) of each link is modeled as the harmonic mean of the actual SNR and $30$ dB. This limits the effective
SNR within $30$ dB. The minimum SNR of a link is assumed to be $0$ dB, i.e., a link is assumed to exist only
if the SNR of the link exceeds $0$ dB. The capacity of each link is obtained using the effective SNR of the link,
total bandwidth and Shannon's capacity theorem.

Table~\ref{tab:LinkBudgetCal} lists the simulation parameters that we have used throughout this paper. 
The next several sub-sections describe simulation results illustrating the advantages and some features
of IAB networks.

\subsection{Comparison between IAB and access only network}
\begin{figure}[!t]
\centering
\includegraphics[width=2.1in]{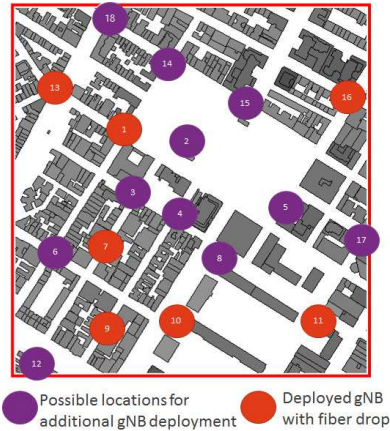}
\caption{The same wireless network of Fig.~\ref{fig:BS_Location_Without_BS} where gNBs with fiber have been deployed 
in $7$ out of the $18$ possible deployment locations.}
\label{fig:BS_Location_No_IAB_7Drop}
\end{figure}
We first select a set of gNBs as anchor nodes, i.e., as nodes with fiber deployment. 
This can be done randomly or using some metric. In this paper, we select the specific anchor nodes
to minimize the total number of anchor nodes while meeting UEs' demands in an access only network. This part of our work 
was previously addressed in~\cite{Nazmul2} and will be skipped here for brevity. Fig.~\ref{fig:BS_Location_No_IAB_7Drop}
shows the location of the gNBs with fiber deployment. The UE locations are not shown in the figure.
Based on the selected set of fiber deployment, we maximize the geometric mean of UE rates throughout
the network and run different versions of the optimization problem that are described in section~\ref{sec:DiffVersion}.
%At first, we focus on access only network with signal-strength-based cell selection and
%IAB network with signal-strength-
%
%IAB winetwork and access only network without 
%any load balancing in access. 
%
\begin{figure}[!t]
\centering
\includegraphics[width=2.1in]{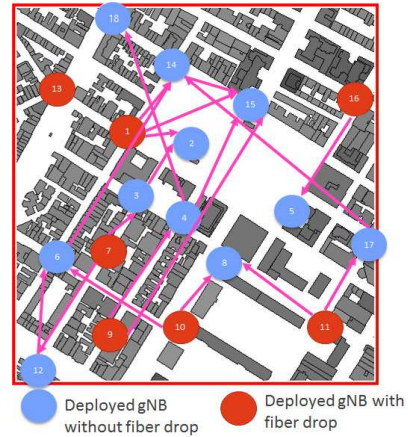}
\caption{
%DL backhaul flow in the network where gNBs have been deployed with and without fiber in $7$ and remaining $11$ out of $18$ possible site locations respectively. 
Routing path of DL backhaul flow in the IAB network with signal-strength-based cell selection and load-balanced mesh topology} 
%The wireless network where gNBs with fiber have been deployed in $7$ out of the $18$ possible deployment locations. In the remaining $11$ locations, gNB have been deployed without fiber. The routing paths among gNBs lead to a mesh network and is designed to maximize the geometric mean of all UEs' rates.
\label{fig:BS_Location_Optimal_Mesh}
\end{figure}
Fig.~\ref{fig:BS_Location_Optimal_Mesh} shows the routing path for downlink flows between gNBs with fiber drop and
gNBs without fiber drop in the IAB network with signal-strength-based cell selection and load-balanced mesh topology. The connectivity between UEs and gNBs are not shown 
in the figures throughout the paper. 
\begin{figure}[!t]
\centering
\includegraphics[width=2.1in]{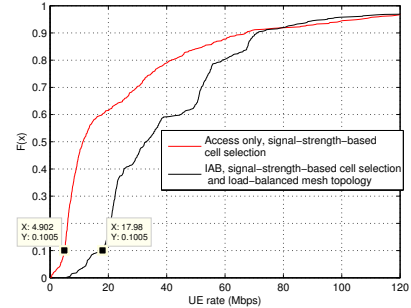}
\caption{UE rate distribution with IAB and access only networks 
with $7$ fiber drops.}
\label{fig:IAB_vs_NoIAB_No_LB}
\end{figure}
Fig.~\ref{fig:IAB_vs_NoIAB_No_LB} shows the distribution of UE rates in both IAB networks and access only networks. 
These results were obtained using signal-strength-based cell selection.
Fig.~\ref{fig:IAB_vs_NoIAB_No_LB} shows that the top $10$ percentile UEs obtain similar rates in both IAB and access only
networks. However, IAB significantly increases the rate of bottom $90$ percentile UEs. 

This result can be intuitively
explained as follows: if cells are selected based on signal strength, 
UEs that are located close to gNBs with fiber drops will get similar rates in both IAB and access only networks.
However, in access only networks, only $7$ gNBs out of $18$ possible site locations are connected to fiber drops.
A lot of UEs get poor rates since they have to be attached with gNBs that are far from them. IAB, on the other hand, brings
gNBs closer to the UEs while ensuring that fiber deployment cost remains the same. Hence, IAB allows network
to significantly increase the rates of these `far' UEs while retaining the rates of the `near' ones.
\begin{figure}[!t]
\centering
\includegraphics[width=2.1in]{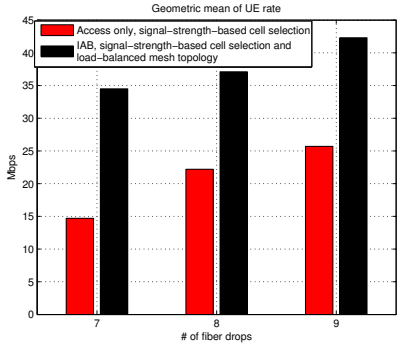}
\caption{Geometric mean of UE rates in IAB and access only networks for different number of fiber drops}
\label{fig:CompGeometricMean}
\end{figure}
Fig.~\ref{fig:CompGeometricMean} shows the geometric mean of UE rates for different number of fiber drops. 
In the considered network setting, as long as
the number of fiber drops is less than the total number of possible site locations, the geometric mean of UE rates 
in IAB networks remains greater than that of access only networks almost by a factor of $2$. 
%After comparing the performance of IAB networks with access only networks, we now investigate 
%the significance of IAB in incremental fiber deployment.
%\centering
%\includegraphics[width=2.7in]{./Pictures/EPS_Pictures/IAB_vs_NoIAB_With_Load_Balancing}
%\caption{UE rate distribution with IAB (generated using optimal Mesh) and no IAB with $7$ fiber drop in the network. Results are obtained with load balancing in access, i.e., UE to BS association is based on maximum signal strength}
%\label{fig:IAB_vs_NoIAB_With_LB}
%\end{figure}
%
\subsection{Illustration of IAB's role in incremental fiber deployment}
%
%\begin{figure}[!t]
%\centering
%\includegraphics[width=2.7in]{./Pictures/EPS_Pictures/BS_Location_NoIAB_18Drop}
%\caption{The wireless network where gNBs with fiber have been deployed in all of the $18$ possible deployment locations.}
%\label{fig:BS_Location_No_IAB_18Drop}
%\end{figure}
%
%
\begin{figure}[!t]
\centering
\includegraphics[width=2.1in]{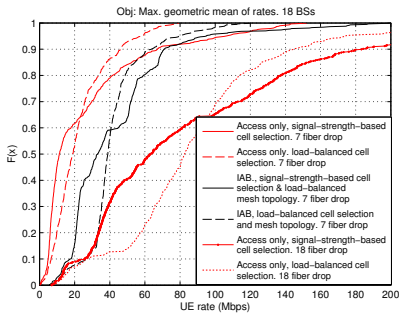}
\caption{Significance of IAB's role during incremental deployment in terms of UE rate}
\label{fig:IAB_NoIAB_7_18Drop}
\end{figure}
\begin{table}
\begin{center}
\begin{tabular}{|l|l|} \hline
Approach & GM of rate \\ \hline
& (Mbps) \\ \hline
Access only, signal-strength-based cell selection ($7$ fiber drops) & $14.7$   \\ \hline
Access only, load-balanced cell selection ($7$ fiber drops) & $17.5$  \\ \hline
IAB, signal-strength-based cell selection and load-balanced  & $34.5$  \\ 
mesh topology ($7$ fiber drops) & \\ \hline
IAB, load-balanced cell selection and mesh topology ($7$ fiber drops) & $38.6$  \\ \hline
Access only, signal-strength-based cell selection ($18$ fiber drops) & $64$ \\ \hline
Access only, load-balanced cell selection ($18$ fiber drops) & $75$  \\ \hline 
\end{tabular}
\end{center}
\caption{Geometric mean of UE rates in three scenarios: 1) access only network with $7$ fiber drops, 2) IAB
mesh network with $7$ fiber drops and 3) access only network with $18$ fiber drops.} \label{tab:GeometricMean_NoIAB_IAB_7_10Drop.}
\end{table}
Fig.~\ref{fig:IAB_NoIAB_7_18Drop} and Table~\ref{tab:GeometricMean_NoIAB_IAB_7_10Drop.} show
IAB's role in incremental fiber deployment.
% by focusing on three scenarios: 1) No IAB with $7$ fiber drops, 2) IAB
%with $7$ fiber drops and 3) no IAB with $18$ fiber drops. 
Results are shown for both signal-strength-based cell selection and load-balanced cell selection.
As mentioned before, compared to access only networks having same number of fiber drops,
IAB almost doubles geometric mean of UE rates.
Deploying fiber to all site locations increases 
geometric mean of UE rates by another factor of $2$. This shows that IAB
can play a significant role during incremental fiber deployment.
\subsection{Comparison between IAB with load-balanced mesh topology and IAB with signal-strength-based spanning tree topology}
\begin{figure}[!t]
\centering
\includegraphics[width=2.1in]{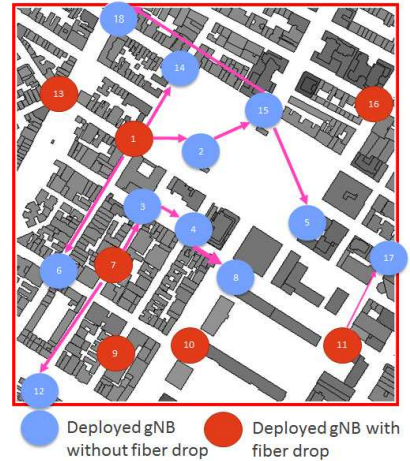}
\caption{Routing path of DL backhaul flow in the IAB network with signal-strength-based cell selection and ST topology}
\label{fig:BS_Location_Spanning_Tree}
\end{figure}
We now investigate the performance of IAB networks with signal-strength-based spanning tree topology. 
Fig.~\ref{fig:BS_Location_Spanning_Tree}
shows the routing pattern of the IAB network where the location of
the fiber drop is same as that of Fig.~\ref{fig:BS_Location_No_IAB_7Drop}.
Since the backhaul connectivity pattern follows a spanning tree,
each gNB without fiber is only connected to one gNB with fiber, potentially over multiple hops.
\begin{figure}[!t]
\centering
\includegraphics[width=2.1in]{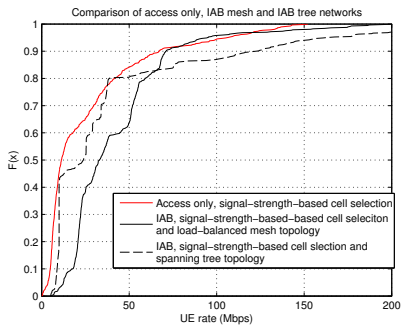}
\caption{UE rate distribution of IAB with load-balanced topology, IAB with signal-strength-based tree topology and access only networks with $7$ fiber drops.}
\label{fig:NoIAB_IAB_ST_Mesh}
\end{figure}
\begin{table}[!t]
\begin{center}
\begin{tabular}{|l|l|} \hline
Approach & GM of rate (Mbps) \\ \hline
Access only, signal-strength-based cell selection   & $14.7$  \\ \hline
IAB, signal-strength-based cell selection \& ST topology & $22.9$  \\ \hline
IAB, signal-strength-based cell selection \& load-balanced   & $34.5$  \\ 
mesh topology & \\ \hline
\end{tabular}
\end{center}
\caption{Comparison of geometric mean of UE rates among three scenarios: 1) Access only, 2) IAB
with load-balanced mesh topology and 3) IAB with signal-strength-based ST topology} \label{tab:GeometricMean_NoIAB_IAB_ST}
\end{table}

Fig.~\ref{fig:NoIAB_IAB_ST_Mesh} and table~\ref{tab:GeometricMean_NoIAB_IAB_ST} compare the performance 
among access only network, IAB with load-balanced mesh topology and IAB with signal-strength-based spanning tree
topology. IAB network with signal-strength based spanning tree topology does not consider access load while
connecting gNBs without fiber points with anchor nodes. As a result, some anchor nodes
get connected to a lot of gNBs and UEs whereas some others get connected to only a few gNBs and UEs.
Hence, the top $20$ percentile UEs enjoy better rates in IAB with signal-strength-based spanning tree topology,
but the bottom $80$ percentile UEs enjoy better rates in IAB with load-balanced mesh topology.
The geometric mean of UE rates of IAB network with signal-strength-based spanning tree topology lies between
that of access only network and IAB network with load-balanced mesh topology.
\begin{figure}[!t]
\centering
\includegraphics[width=2.1in]{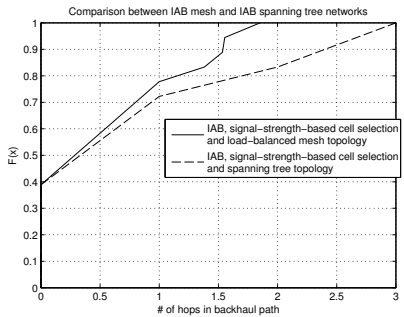}
\caption{Number of hops between gNBs and anchor nodes with IAB based on optimal mesh (generated using the routing pattern of Fig.~\ref{fig:BS_Location_Optimal_Mesh}) and IAB based on spanning tree (generated using the routing pattern of Fig.~\ref{fig:BS_Location_Spanning_Tree})}
\label{fig:HopCountCDF}
\end{figure}

Fig.~\ref{fig:HopCountCDF} shows hop count distribution between gNBs and anchor nodes
in both IAB scenarios. 
As shown in Fig.~\ref{fig:BS_Location_Optimal_Mesh}, a gNB may 
have multiple routes to its anchor(s) in an IAB network with load-balanced mesh tology. The hop count for these gNBs is determined
as the rate-based mean value of the hop counts of its multiple routes.

Since there are $7$ anchor nodes in the simulation setting, the number of hops 
for approximately $40\%$ gNBs ($7$ out of $18$) is zero in both IAB scenarios. Since IAB with signal-strength-based 
spanning tree does not consider access load while generating 
backhaul routing pattern, hop count in this scenario is significantly
greater than that of IAB with optimal mesh networks. Hence, IAB with signal-strength-based
spanning tree will suffer from higher latency as well.

\section{Conclusion}   \label{sec:Conclusion}

Wireless backhaul increases both coverage and capacity in mobile networks
and can play a crucial role during incremental deployment of fiber. 
IAB allows inter-operability among 
base stations from different manufacturers, which is essential
for flexible deployment of dense small cell networks.
This work investigates different aspects of IAB networks. Simulation results suggest
that as long as the number of fiber drops is less than half of the total number of possible site locations,
the geometric mean of UE rates in an IAB network remains almost a factor of $2$ higher than
that in an access only network. Besides,
IAB network with signal strength based spanning tree topology performs
worse than that with load-balanced mesh topology in terms of UE rate and latency.
The performance of an IAB network with spanning tree
can be improved by considering access load while generating the tree. Design and simulation of an optimal IAB network
with access load and signal strength based spanning tree in MMW band remains an area of future work for our study.

%\appendix{Fiber Drop Deployment to Meet Demand}

\bibliographystyle{./IEEEtran}
\bibliography{IABmmSys2018}

%\section{References}
%\input{IABReference}

\end{document}